\documentclass[aps,pra,superscriptaddress,twocolumn]{revtex4-2}

\usepackage[utf8]{inputenc}
\usepackage{amsmath, mleftright}
\usepackage{braket}
\usepackage{amsfonts}
\usepackage{amssymb}
\usepackage{lipsum}
\usepackage{dsfont}
\usepackage{stmaryrd}
\usepackage{graphicx}
\usepackage{booktabs}
\usepackage{siunitx}
\usepackage[version=4]{mhchem}
\usepackage{hyperref}
\usepackage[capitalise]{cleveref}
\usepackage{siunitx}
\usepackage{mathtools}
\usepackage{xcolor}
\usepackage{microtype}
\usepackage{algorithm2e}
\usepackage{algpseudocode}
\usepackage{acro}
\usepackage[normalem]{ulem}

\newcommand{\approptoinn}[2]{\mathrel{\vcenter{
    \offinterlineskip\halign{\hfil$##$\cr
    #1\propto\cr\noalign{\kern2pt}#1\sim\cr\noalign{\kern-2pt}}}}}

\newcommand{\stkout}[1]{\ifmmode\text{\sout{\ensuremath{#1}}}\else\sout{#1}\fi}

\def\conflictsname{Conflicts of interest}
\newenvironment{conflicts}{%
    \expandafter\section\expandafter*\expandafter{\conflictsname}%
    }%

\newcommand{\auchem}{Department of Chemistry, Aarhus University, DK-8000 Aarhus C, Denmark}
\newcommand{\auphys}{Department of Physics and Astronomy, Aarhus University, DK-8000 Aarhus C, Denmark}
\newcommand{\kvantify}{Kvantify Aps, DK-2300 Copenhagen S, Denmark}

\DeclareAcronym{hf}{
    short = HF ,
    long = Hartree-Fock ,
}
\DeclareAcronym{fci}{
    short = FCI ,
    long = full configuration interaction ,
}
\DeclareAcronym{vqe}{
    short = VQE ,
    long = variational quantum eigensolver ,
}
\DeclareAcronym{adapt}{
    short = ADAPT-VQE ,
    long =  Adaptive Derivative-Assembled Pseudo-Trotter Ansatz Variational Quantum Eigensolver ,
}
\DeclareAcronym{algo}{
    short =  FAST-VQE,
    long = Fermionic Adaptive Sampling Theory VQE ,
}
\DeclareAcronym{en}{
    short = EN ,
    long = Epstein-Nesbet ,
}
\DeclareAcronym{enlm}{
    short = HSCI ,
    long = Heuristic Selected CI,
}
\DeclareAcronym{dsgn}{
    short = HG ,
    long = Heuristic Gradient ,
}
\DeclareAcronym{nisq}{
    short = NISQ ,
    long = Noisy Intermediate-Scale Quantum  ,
}
\DeclareAcronym{sci}{
    short = SCI ,
    long = Selected Configuration Interaction,
}
\DeclareAcronym{jw}{
    short = JW ,
    long = Jordan-Wigner ,
}
\DeclareAcronym{qeb_adapt_vqe}{
    short = QEB-ADAPT-VQE ,
    long = Qubit Excitation Based ADAPT-VQE ,
}

\begin{document}
    \title{Fermionic Adaptive Sampling Theory for Variational Quantum Eigensolvers}

    \author{Marco Majland}
    \affiliation{\kvantify}
    \affiliation{\auphys}
    \affiliation{\auchem}

    \author{Patrick Ettenhuber}
    \affiliation{\kvantify}

    \author{Nikolaj Thomas Zinner}
    \affiliation{\kvantify}
    \affiliation{\auphys}

    \begin{abstract}
        Quantum chemistry has been identified as one of the most promising areas where
        quantum computing can have a tremendous impact.
        For current \ac{nisq} devices, one of the best available methods to prepare approximate wave functions
        on quantum computers is the \ac{adapt}. However, ADAPT-VQE suffers from a significant measurement
        overhead when estimating the importance of operators in the wave function. In this work, we propose \ac{algo},
        a method for selecting operators based on importance metrics solely derived from the populations of Slater determinants in the
        wave function.
        Thus, our method mitigates measurement overheads for \ac{adapt} as it is only dependent on the populations of Slater
        determinants which can simply be determined by measurements in the computational basis. We introduce two heuristic
        importance metrics, one based on Selected Configuration Interaction with perturbation theory and one based on
        approximate gradients. In state vector and finite shot simulations, \ac{algo} using the heuristic
        metric based on approximate gradients converges at the same rate or faster than \ac{adapt}
        and requires dramatically fewer shots.

    \end{abstract}

    \maketitle

    \section{Introduction}\label{sec:introduction}

    Quantum chemistry has been identified as one of the most promising areas where
    quantum computing can have great impact on industrial applications\cite{elfving_how_2020,daley_practical_2022,gonthier_identifying_2020,aspuru-guzik_simulated_2005}.
    However, current quantum hardware is subject to noise and error and thus algorithms such as quantum phase
    estimation remain intractable for current and near-term devices \cite{bharti_noisy_2022,abrams_quantum_1999}. Therefore, the research
    community has focused on developing algorithms suitable for an era of noise, error, limited qubits and limited quantum gates~\cite{mcclean_theory_2016,kandala_hardware-efficient_2017}.
    A promising method to approximate electronic wave functions on quantum computers is the \ac{adapt} algorithm,
    along with its variants, which has made tremendous progress towards this goal\cite{grimsley_adaptive_2019,tang_qubit-adapt-vqe_2021,yordanov_qubit-excitation-based_2021,lan_amplitude_2022,anastasiou_tetris-adapt-vqe_2022,bertels_symmetry_2022}.
    Other adaptive algorithms include the Qubit Coupled Cluster method and the Iterative Qubit Coupled Cluster method~\cite{ryabinkin_qubit_2018,ryabinkin_iterative_2019}.
    The adaptive approaches for estimating electronic wave functions contrast the static approaches such as Unitary Coupled
    Cluster Theory and its variants\cite{romero_strategies_2018,anand_quantum_2022,lee_generalized_2019}.\\

    The adaptive algorithms have proven to converge to chemical accuracy with fewer parameters and more compact wave functions
    compared to that of the static algorithms. Thus, the adaptive algorithms may be more feasible for near-term applications.
    However, one of the primary challenges of \ac{adapt} is the large measurement overhead incurred by estimating the importance metric
    for selecting relevant operators for the wave function~\cite{grimsley_adaptive_2019}. Even estimating a single energy
    evaluation of a wave function through the sampling of expectation values may require significant measurement resources
    as was demonstrated in recent large-scale benchmarks~\cite{gonthier_identifying_2020}. For \ac{adapt}, the importance
    metric for choosing operators from a predefined pool, $\mathcal{A}$, is the gradient of the energy. Therefore,
    the number of measurements necessary to rank the operators scales with the size of the pool, i.e. $\mathcal{O}(|\mathcal{A}|)$.
    Since $\mathcal{A}$ typically contains two-body operators, the size of the set of operators $|\mathcal{A}|$ scales as $\mathcal{O}(N^4)$,
    where $N$ is a measure for the size of the chemical system. \\

    In this work, we propose a method for selecting operators based on the populations of
    Slater determinants in the wave function in order to establish an importance metric for excitation
    operators. This is in stark contrast to \ac{adapt} where the importance of operators is established using gradient
    measurements which requires the sampling of expectation values for each excitation operator.
    Sampling Slater determinants requires only the sampling of a single operator rather than $\mathcal{O}(N^{4})$ operators as
    in \ac{adapt}. In fact, the required quantities for evaluating the proposed metric can be extracted from a measurement of
    the energy in \ac{vqe}, a measurement that would in any case have to be performed.\\

    For selecting operators, we are considering two metrics, one that is related to the approximate gradient used in \ac{adapt} and
    a second one that is inspired by classical \ac{sci}\cite{huron_iterative_1973}. In classical \ac{sci}, the determinants used to diagonalize the Hamiltonian are chosen using an importance metric
    typically based on a perturbation method\cite{bytautas_priori_2009,anderson_breaking_2018,bender_studies_1969,whitten_configuration_1969,evangelisti_convergence_1983}.
    Here we consider selecting operators based on second-order \ac{en} perturbation theory\cite{epstein_stark_1926,nesbet_configuration_1997}.
    The methods are compared to \ac{adapt} by calculating the ground state energies of two small molecules which
    are typically used in benchmarks, namely $\text{H}_{4}$ and $\text{LiH}$. The ground state energies are calculated using
    state vector (infinite shot) and finite shot simulations to investigate the performance of the methods both cases. 
		
		The paper is organized as follows. In Sec.~\ref{sec:background}, we provide the theoretical background
    of \ac{adapt} and \ac{sci}. In Sec.~\ref{sec:algo} we provide the background for the scaling reduction in \ac{algo}
    and derive the gradient-based and \ac{sci}-based metrics.
    In Sec.~\ref{sec:compdetails}, we
    provide a pseudo-algorithm for \ac{algo} and provide the computational details of our calculations which
    we will present and discuss in Sec.~\ref{sec:results}. Finally, we 
    conclude with a summary and present some future research avenues in Sec.~\ref{sec:conclusion}.

    \section{Background} \label{sec:background}
    In this section, we will provide the background necessary for understanding the construction
    of our method in Sec.~\ref{sec:algo}, starting with \ac{adapt} and followed by \ac{sci}.

    \subsection{ADAPT-VQE}
    In \ac{adapt}, an Ansatz is built by successively adding parametrized unitary operators
    acting on a reference state $\ket{\Phi_0}$, which is often taken as the \ac{hf}
    ground state determinant. Thus, the \ac{adapt} wave function in iteration $k$ of the
    algorithm can be expressed as
    \begin{equation}
        \ket{\Psi^{(k)}} = \prod_{\mu\in\mathcal{A}^{(k)}} e^{-i\theta_\mu\hat{A}_\mu} \ket{\Phi_0}, \label{eq:expans}
    \end{equation}
    where $\mathcal{A}^{(k)}$ is the set of operators in the wave function at iteration $k$,
    $\hat{A}_\mu = \hat{\tau}_\mu - \hat{\tau}^\dagger_\mu$, with $\hat{\tau}_\mu$ being an excitation operator and $\mu$
    enumerates the excitation. The excitation operators are chosen from a pool of operators, $\mathcal{A}=\{A_{\mu}\}$, based on
    an importance metric, $w(\hat{A}_\mu, \ket{\Psi^{(k)}})$.
    In standard \ac{adapt}, the importance metric is the gradient of the energy with respect to the parameter of the operator.
    The energy of the $k+1$st iteration may be written as
    \begin{equation}
        E^{(k+1)} = \langle\Psi^{(k)} | e^{i\theta_\mu \hat{A}_\mu} \hat{H} e^{-i\theta_\mu \hat{A}_\mu} |\Psi^{(k)} \rangle
    \end{equation}
    such that
    \begin{equation}
        \begin{split}
            g_\mu &= \mleft. \frac{\partial E^{(k+1)}}{\partial \theta_\mu}\mright|_{\theta_\mu = 0} \\
            &= i \langle\Psi^{(k)} | [\hat{A}_\mu, \hat{H}] | \Psi^{(k)} \rangle. \label{eq:adaptgrad}
        \end{split}
    \end{equation}
    To evaluate this expression, \ac{adapt} relies on measuring operators of the type $[\hat{A}_\mu, \hat{H}]$,
    yielding a significant overhead in measurements to be performed.

    \subsection{Selected CI}
    In \ac{sci}, determinants are selected iteratively by an importance metric in order to adaptively increase the subspace in which the CI eigenvalue problem is solved.
    One possibility for selecting determinants is based on perturbation theory\cite{huron_iterative_1973}. In this paper, we consider \ac{en}
    perturbation theory\cite{epstein_stark_1926,nesbet_configuration_1997}. \ac{en} theory weights the importance of a Slater determinant $|D\rangle$ for extending a
    wave function $|\Psi^{(k)}\rangle$ in iteration $k$ as
    \begin{equation}
        \begin{split}
            E^{(k)}_{D} &= \frac{|\langle D | \hat{H} | \Psi^{(k)} \rangle|^2}{E^{(k)} - \langle D | \hat{H} | D \rangle}\\
            &= \sum_{ij}\frac{c_i c_j^* \langle D_j | \hat{H} | D\rangle\langle D | \hat{H} | D_i\rangle}{E^{(k)} - \langle D | \hat{H} | D \rangle}, \label{eq:en}\\
        \end{split}
    \end{equation}
    where the states $|D_i\rangle$ are Slater determinants and $c_i = \langle D_i | \Psi^{(k)} \rangle$ CI coefficients.

    \section{FAST-VQE} \label{sec:algo}
    In this section, we present a method for selecting operators solely based on the population
    of Slater determinants in the wave function by establishing importance metrics for excitation
    operators. This is in stark contrast to \ac{adapt} where the importance of operators is established by measuring
    the expectation value of the non-diagonal gradient operators of Eq.~\eqref{eq:adaptgrad}. We start this section
    with a discussion of sampling populations of Slater determinants and diagonal Hamiltonian measurements in Sec.~\ref{sec:diagh}
    and then build the two metrics in Secs.~\ref{sec:enlm} and \ref{sec:dsgn}.

    \subsection{Sampling populations of Slater determinants} \label{sec:diagh}
    A population of Slater determinants may, for example, be obtained from the energy evaluations in the \ac{vqe} optimization
    or as a separate measurement. For separate measurements, given $\ket{\Psi^{(k)}}$, one may repeatedly perform measurements
    in the computational basis to obtain a bit string representation of determinants from $\ket{\Psi^{(k)}}$ in the \ac{hf} basis.
    These measurements may be collected in a multi-set of determinants. The multi-set may be written as
    \begin{equation}
        S^{(k)} = \left\{ |D_i\rangle , \langle D_i | \Psi^{(k)} \rangle \neq 0\right\}, \label{eq:multiset}
    \end{equation}
    where the frequency of each determinant $|D_i\rangle$ is proportional to $|c_i|^2$ and where the restriction
    is fulfilled by construction. With this set of determinants, we can build metrics suitable to assign
    importance weights to operators from an operator pool $\mathcal{A}$ based on the expected contribution to the
    wave function. In the following sections we will introduce two such metrics.\\
    For energy measurements, the population of Slater determinants may be obtained through
    sampling the diagonal elements of the Hamiltonian. In \ac{vqe}, the Hamiltonian is mapped to
    a qubit Hamiltonian,
    \begin{equation}
        \hat{H} = \sum_{a} h_{a} \hat{P}_{a},
    \end{equation}
    where
    \begin{equation}
        \hat{P}_{a} = \bigotimes_{b} \hat{\sigma}_{b}^{\alpha}, \quad \alpha\in\{x,y,z\},
    \end{equation}
    denotes a product of Pauli operators. Consider a partitioning of the Hamiltonian $\hat{H} = \hat{H}^z + \hat{H}^c$ where $\hat{H}^z$ is diagonal, then
    we can express $\hat{H}^z$ as $\hat{H}^z = \sum_{a} h_{a} \hat{P}^z_{a}$, where $\hat{P}^z_{a}$ are products of Pauli-$z$ operators.
    We can then write an energy functional depending on the wave function parameters $\boldsymbol{\theta}$ in terms of this partitioned Hamiltonian as
    \begin{equation}
        \begin{split}
            E^{(k)}(\boldsymbol{\theta}) &= \langle \Psi^{(k)}| \hat{H}^z + \hat{H}^c |\Psi^{(k)}\rangle \\
            & = \sum_{a} h_{a} \langle \Psi^{(k)}| \hat{P}^z_a |\Psi^{(k)}\rangle + \langle \Psi^{(k)}| \hat{H}^c |\Psi^{(k)}\rangle\\
            & = \sum_{ai} h_{a} |c_i|^2 \langle D_i | \hat{P}^z_a |D_i\rangle + \langle \Psi^{(k)}| \hat{H}^c |\Psi^{(k)}\rangle \\
            & = \sum_{i} h_{ii} |c_i|^2 + \langle \Psi^{(k)}| \hat{H}^c |\Psi^{(k)}\rangle. \label{eq:equation2}
        \end{split}
    \end{equation}
    Thus, we can perform measurements of diagonal Hamiltonian terms in the computational basis in order to sample
    Slater determinants $|D_i\rangle$ in $|\Psi^{(k)}\rangle$ with a probability that is proportional to $|c_i|^2$. 
    Note that Eq.~\eqref{eq:equation2} is evaluated repeatedly in order to optimize the wave function
    parameters $\boldsymbol{\theta}$, e.g. using \ac{vqe}, such that no additional cost is introduced to calculate
    Slater determinant populations.

    \subsection{Heuristic Gradient} \label{sec:dsgn}
    To introduce the first heuristic importance metric, we start from $g_\mu$ in Eq.\eqref{eq:adaptgrad} which may be expressed as
    \begin{equation}
        \begin{split}
            g_\mu &= i \langle\Psi^{(k)} | \hat{A}_\mu \hat{H} - \hat{H}\hat{A}_\mu  | \Psi^{(k)} \rangle\\
            &= -i \langle\Psi^{(k)} | \hat{A}^\dagger_\mu \hat{H} + \hat{H}\hat{A}_\mu  | \Psi^{(k)} \rangle \\
            &= -2i \Re(\langle \Psi^{(k)} | \hat{A}^\dagger_\mu \hat{H} | \Psi^{(k)}\rangle)\\
            &= -2i \sum_{ij} \Re(c_i^*c_j\langle D_i | \hat{A}^\dagger_\mu \hat{H} | D_j\rangle). \label{eq:altgrad}\\
        \end{split}
    \end{equation}
    Then, dropping the off-diagonal part of the sum in Eq.~\eqref{eq:altgrad} yields
    \begin{equation}
        \begin{split}
            \text{diag}(g_\mu) &= 2i \sum_i \Re( |c_i|^2 \langle D_i | \hat{A}^\dagger_\mu \hat{H} | D_i \rangle) \\
            &= 2i \sum_i |c_i|^2 \Re(\langle D_i | \hat{A}^\dagger_\mu \hat{H} | D_i \rangle). \label{eq:exaltmeasure}
        \end{split}
    \end{equation}
    The manifold into which $\hat{A}^\dagger_\mu$ excites, $\{\langle D_j|\hat{A}^\dagger_\mu, D_j \in S^{(k)}\}$,
    may be classically constructed. Such a manifold contains information on how the diagonal is connected to off-diagonal
    elements. To include that information in the final metric, a second sum over the determinants will therefore be introduced.
    In this regard, $S^{(k)}$ of Eq.~\eqref{eq:multiset} will be used directly since the number of occurrences of a determinant $\ket{D_{i}}$
    in this multiset is proportional to $|c_i|^2$. Additionally, the second summation is introduced and all prefactors are removed,
    as the final ranking will not be dependent on constant factors. Thus, one obtains
    \begin{equation}
        \alpha_\mu = \sum_{D_i \in S^{(k)}}\sum_{D_j \in S^{(k)}} \Re(\langle D_i | \hat{A}^\dagger_\mu \hat{H} | D_j \rangle), \label{eq:altmeasure}
    \end{equation}
    which concludes the construction of the first importance metric. This metric roughly corresponds to dropping the
    phases and prefactors from Eq.~\ref{eq:altgrad}. Note that this expression can be evaluated classically once
    $S^{(k)}$ has been obtained. This importance metric will be denoted \ac{dsgn} in the following.\\
    In contrast to \ac{adapt}, it is necessary to remove operators already used in the Ansatz, $\mathcal{A}^{k}$,
    from the operator pool, $\mathcal{A}$, in order to avoid using the same operator twice. However, to converge to the \ac{fci}
    energy, it may be necessary to repeat operators in the Ansatz. Thus, whenever $\max_{\mu}(\alpha_\mu)<\epsilon$,
    the operators $\mathcal{A}^{k}$ are added to the pool again.

    \subsection{Heuristic Selected CI} \label{sec:enlm}
    In order to introduce a second heuristically motivated metric, \ac{sci} theory will be leveraged.
    In contrast to \ac{sci} theory, which works with the determinants directly, it is required to build a metric that
    relates determinants and their frequencies in the sampling procedure to operators in order to gauge the effect of adding
    them to the Ansatz. In this section, such a metric will be constructed based on the \ac{en} criterion
    from Eq.~\eqref{eq:en}. 
		
		First, consider the \ac{adapt} Ansatz in Eq.~\eqref{eq:expans}. The addition of a new
    operator corresponds to the multiplication of a new exponential
    which operates on all previous exponentials and the reference wave function. Thus, the contribution must be
    evaluated for all determinants already in $|\Psi^{(k)}\rangle$ and appropriately weighted. The construction of the
    heuristic operator metric based on determinants begins by noting that
    $\langle D_i | \hat{A}_\mu^\dagger = \langle D_k |$ is just another determinant or zero, establishing
    a connection between operators and determinants. From this, it would be possible to evaluate
    the contribution of $\hat{A}_\mu^\dagger$ by applying the \ac{en} criterion in Eq.~\eqref{eq:en} directly using $D_j$ as
    the contribution to be evaluated. However, naively sampling the operator $H\ket{D_k}\bra{D_k}H$ comes at a significant cost with a scaling
    of $\mathcal{O}(N^{8})$. In order to make this manageable and to be able to evaluate this on a classical computer, the
    off-diagonal elements of the sum over $i$ and $j$ from Eq.~\eqref{eq:en} may be neglected. Furthermore, one must evaluate
    and sum such a metric for all the determinants an operator $\hat{A}_\mu^\dagger$ is able to create from the determinants
    in $|\Psi^{(k)}\rangle$, i.e., for practical implementations, all the determinants of the multi-set $S^{(k)}$.
    For representing the wave function in Eq.~\eqref{eq:en}, the same approach as used to arrive at Eq.~\eqref{eq:altmeasure}
    will be used, i.e., a finite shot representation given the determinants collected in $S^{(k)}$ and using only the diagonal
    contributions.
		
    This concludes the construction of the heuristic importance metric $\beta_\mu$, which may be written as
    \begin{equation}
        \begin{split}
            \beta_\mu & \coloneqq  \sum_{D_i\in \mathcal{S}^{(k)}} \sum_{D_j\in \mathcal{S}^{(k)}} \frac{|\langle D_i | \hat{A}^\dagger_\mu \hat{H} | D_j \rangle|^2}{E^{(k)} - \langle D_i | \hat{A}^\dagger_\mu\hat{H} \hat{A}_\mu | D_i\rangle}. \label{eq:weight}
        \end{split}
    \end{equation}
    This importance metric will be denoted as \ac{enlm}. Note that also for this metric, we need to remove used operators
    from $\mathcal{A}$, as explained in Sec.~\ref{sec:dsgn}.\\
    The importance metrics in Eqns.~\eqref{eq:altmeasure} and~\eqref{eq:weight} both use the operator $\hat{A}^\dagger\hat{H}$ for the evaluation
    of the importance of an operator $\hat{A}_\mu$ when improving the wavefunction in the next iteration. From a set of determinants, it is trivial to evaluate
    the expectation values for this operator on a classical computational resource with polynomial scaling in the number of electrons and orbitals.

    \section{Computational details} \label{sec:compdetails}
    \begin{figure*}[ht]
        \includegraphics[width=\textwidth]{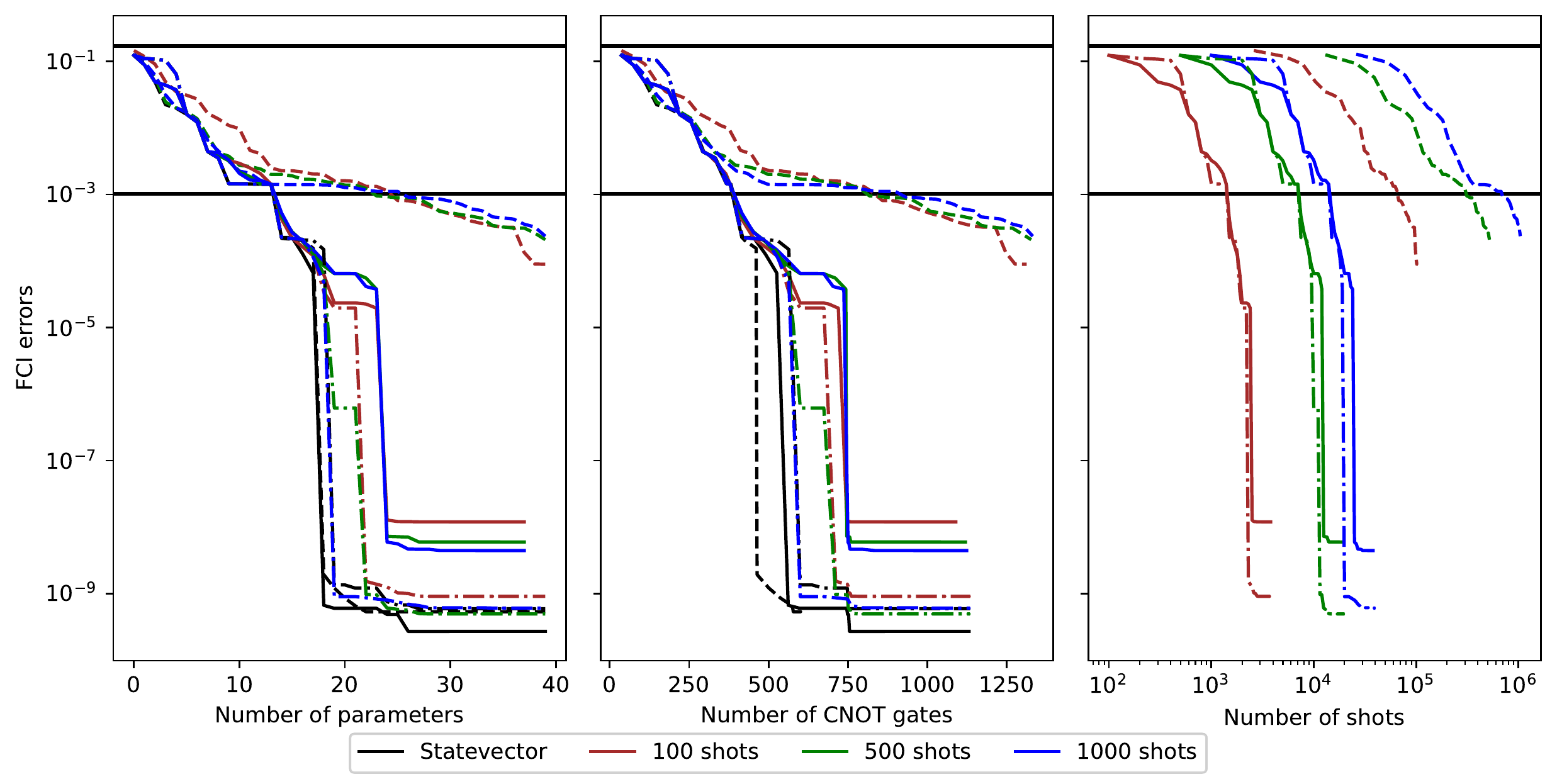}
	    \caption{Convergence of \ac{adapt} (dashed), \ac{enlm} (full)  and \ac{dsgn} (dotted) with respect to the number of parameters (left),
            CNOT gates (middle) and the total number of shots (right) at a given precision of the obtained wave function
            for a linear $\text{H}_4$ chain at 1.5 Angstrom separation between subsequent atoms.}
        \label{fig:h4}
    \end{figure*}

    In this section, the algorithms and computational details of the calculations will be reviewed, starting with a description of the algorithm in
    Sec.~\ref{sec:algodesc}, a description of the choice of operator pool in Sec.~\ref{sec:opchoice} and finally with a description of the numerical
    experiments in Sec.~\ref{sec:molecules}

    \subsubsection{Review of algorithms}\label{sec:algodesc}
    The general
    algorithm for \ac{adapt} and \ac{algo} is presented in Alg.~1. Note that the major difference between these methods
    is the skipping of lines 7-10 for \ac{adapt}. For \ac{adapt}, the importance metric reads $w(\hat{A}_\mu,\ket{\Psi^{k}})_\textrm{ADAPT-VQE} = g_\mu$,
    while for \ac{algo} we are using the importance metrics introduced earlier, i.e., 
    $w(\hat{A}_\mu,~|\Psi^{(k)}\rangle)_\textrm{HG}~=~\alpha_\mu$
		and~$w(\hat{A}_\mu,~|\Psi^{(k)}\rangle)_\textrm{HSCI}~=~\beta_\mu$.

    Note that modifications for \ac{adapt}, for example TETRIS-ADAPT-VQE\cite{anastasiou_tetris-adapt-vqe_2022},
    which adds more than one operator per iteration, are also applicable to \ac{algo}. However, we do not
    expect the relative performance of the algorithms to differ when using these types of improvements since
    the importance metrics are identical for the operators despite adding more than one operator per iteration.
    Thus, standard implementations for \ac{adapt} and \ac{algo} are used. 
    \vspace*{0.5cm}
    \includegraphics[scale=0.9]{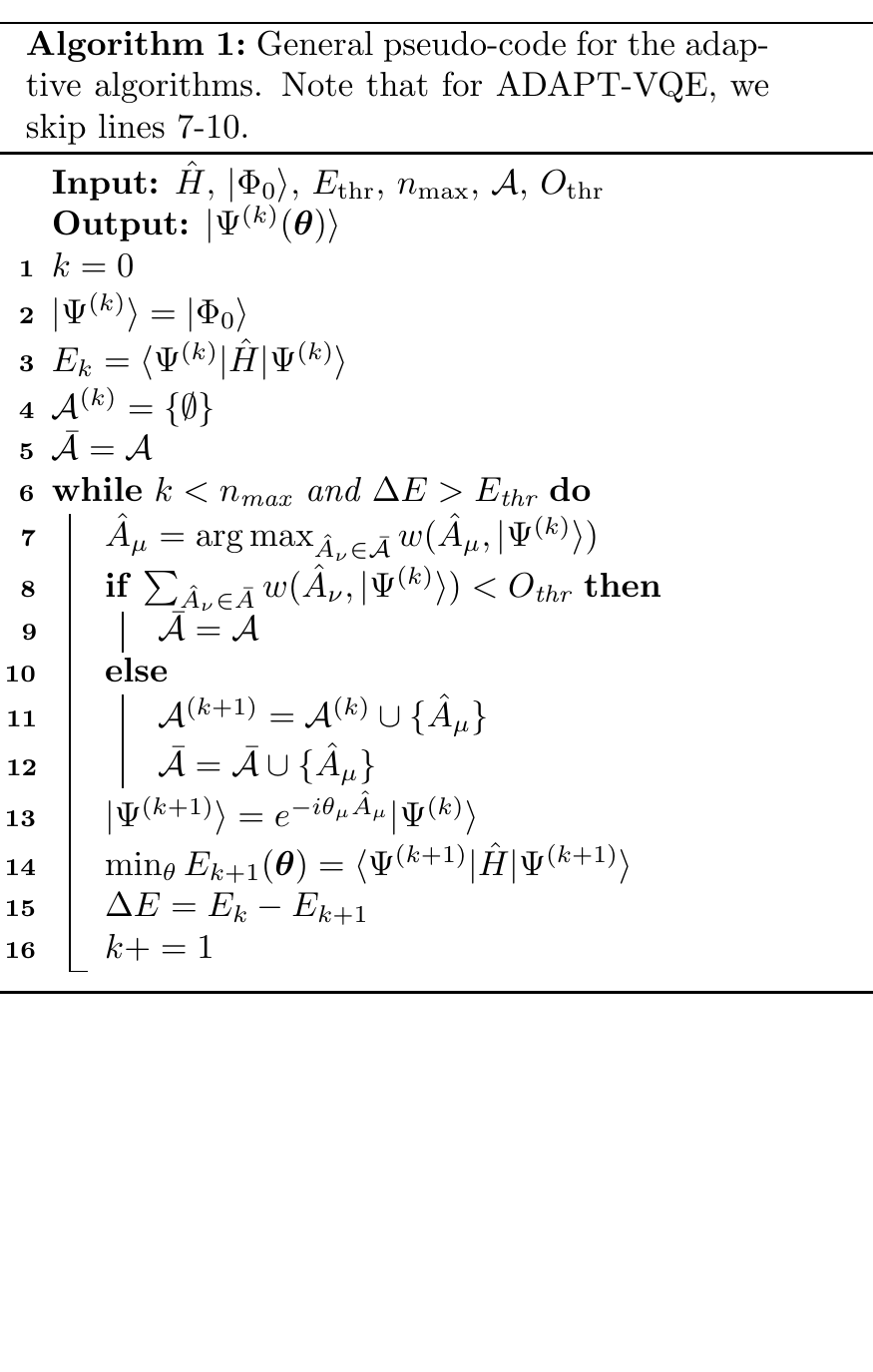}
    \vspace*{-3.5cm}

    \subsubsection{Choice of operator pools}\label{sec:opchoice}
    In general, any type of operator pool may be utilized. However, one-body and two-body
    excitation operators are enough to parametrize an FCI wave function~\cite{evangelista_exact_2019}.
    Since the quantum gates required for implementing $N$-body excitation operators increase rapidly with $N$, operator pools are
    typically restricted to one-body and two-body excitation operators. According to Ref.~\cite{evangelista_exact_2019}
    all possible many-body operators may be decomposed as one-body and two-body excitation operators, specifically as
    infinite sequences of one- and two-body particle-hole operators. Particle-hole excitation operators
    are excitation operators which annihilate electrons in occupied spin-orbitals in the \ac{hf} reference state and create
    electrons in virtual spin-orbitals of the \ac{hf} reference.
    In the original formulation of \ac{adapt}, the operator pool consisted of general excitations (particle-hole excitations
    and excitations within the pure virtual-virtual
    or occupied-occupied blocks) in the \ac{jw}
    encoding\cite{jordan_uber_1928}. The resulting operator pools determine the scaling and convergence of the procedures.
    Additionally, rather than using these physically motivated operator pools, one can build operator pools that are
    computationally motivated. For example, several approximations have been suggested such as \ac{qeb_adapt_vqe}~\cite{yordanov_qubit-excitation-based_2021} and
    spin-adapted ADAPT-VQE~\cite{tsuchimochi_adaptive_2022}. Recently, operator pools which consider qubit-space operators
    were suggested~\cite{tang_qubit-adapt-vqe_2021}.
    In this article, we will use particle-hole excitation operators in
    the \ac{qeb_adapt_vqe} encoding since the primary task of this paper is to investigate importance metrics and not the
    operators themselves.

    \subsubsection{Systems and details}\label{sec:molecules}
    Benchmarks of the algorithms are performed by calculating the ground state energy for $\text{H}_4$ and $\text{LiH}$
    which are typically used to benchmark \ac{adapt} algorithms \cite{grimsley_adaptive_2019,romero_strategies_2018,ryabinkin_iterative_2019}.
    In these calculations, the STO-3G basis set were used.
    The molecular integrals were obtained using PySCF. The optimization of the wavefunction parameters in \ac{vqe}
    is calculated with the L-BFGS-B method as implemented in Qiskit~\cite{Qiskit}. For all molecules, four types of
    calculations were performed, one state vector simulation and three simulations
    with finite sampling (100, 500 and 1000 shots per expectation value estimation). The optimization of the wave function
    in the \ac{vqe} was performed using statevector simulations since we are restricting our study to the evaluation of importance
    measures for operators. Thus, the method for re-using \ac{vqe} optimization measurements for \ac{algo} was not used
    such that finite shot simulations were performed to estimate population of Slater determinants. Since the identical
    number of operators must be sampled in the \ac{vqe} optimization for each algorithm, we do not expect the relative
    comparison between the \ac{adapt} and \ac{algo} to differ in terms of \ac{vqe} optimization.  The state vector
    and finite shot calculations were performed in Qiskit. The state vector simulation serves as a benchmark for
    infinite shots. All quantum simulations are compared to an FCI calculation for the same molecule/basis set
    combination in PySCF. These results are presented in Sec.~\ref{sec:results}.

    \section{Results} \label{sec:results}
    \begin{figure*}[ht]
        \includegraphics[width=\textwidth]{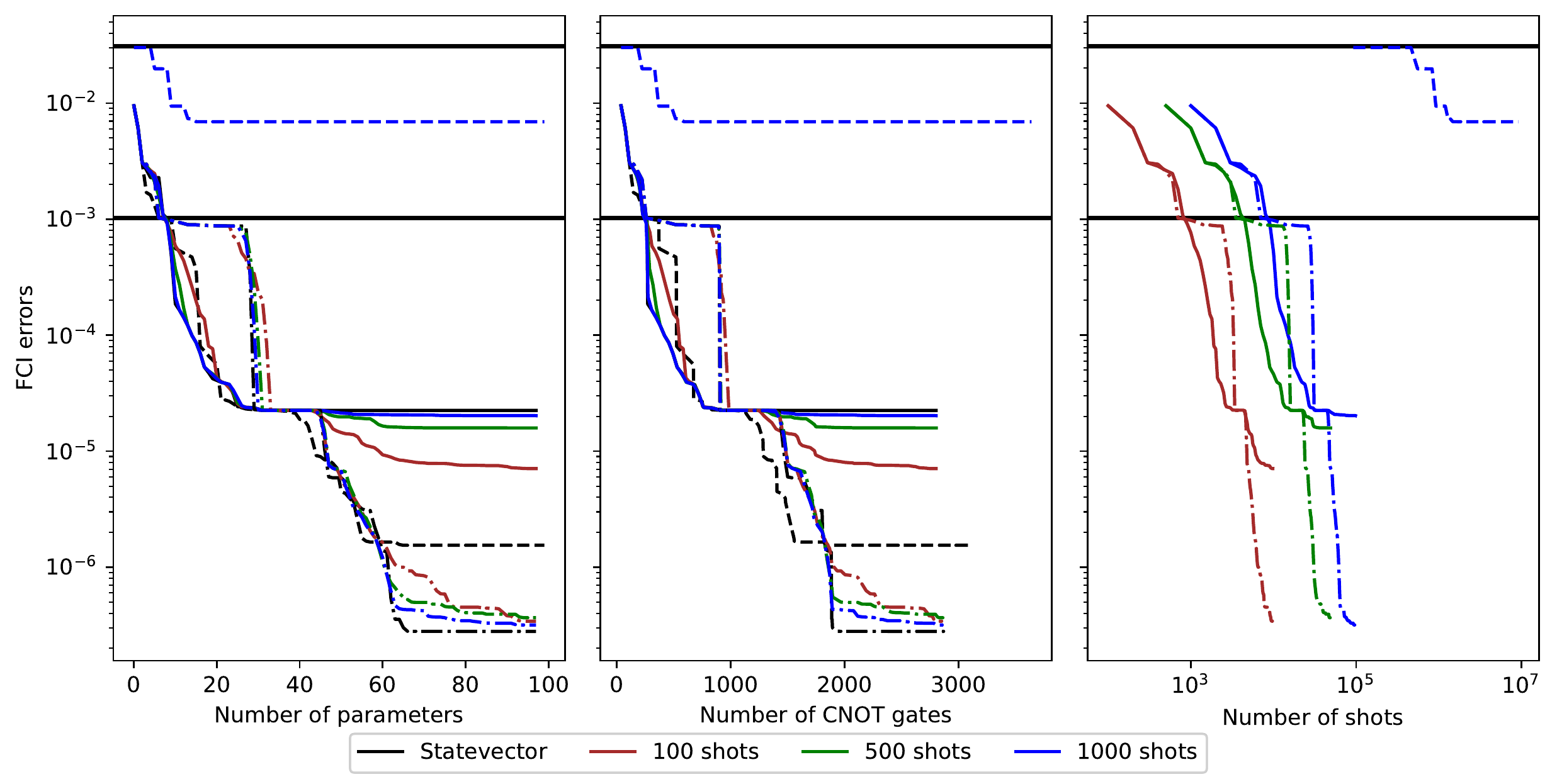}
	    \caption{Convergence of \ac{adapt} (dashed), \ac{enlm} (full) and \ac{dsgn} (dotted) with respect to the number of parameters (left),
            CNOT gates (middle) and the total number of shots (right) at a given precision of the obtained wave function
            for a linear $LiH$ molecule with 1.5 Angstrom separation between the atoms. The numbers for the finite shot simulations for
	    \ac{adapt} are not shown as the energy remained on the \ac{hf} level.}
        \label{fig:lih}
    \end{figure*}
    In this section, the results from the setup described in Sec.~\ref{sec:compdetails} are presented. We will conclude this section
    with a discussion of the results.

    \subsection{H4}
    In Fig.~\ref{fig:h4}, we present the ground state calculations using \ac{adapt} and \ac{algo} with both importance
    metrics, \ac{enlm} and \ac{dsgn}, for a linear
    chain of $\text{H}_4$ in terms of the error relative to the FCI ground state energy. To an error of above $10^{-3}$ Hartree
    with respect to FCI, the convergence in terms of the number of operators (parameters) added to the Ansatz is very similar 
    for all methods and numbers of shots per operator evaluation.
    Beyond that point, the fastest convergence is observed for the state vector simulation for \ac{adapt} closely followed by
    the state vector for \ac{enlm} and \ac{dsgn} and finite shot simulations for \ac{dsgn}. \ac{enlm} converges slower
    for finite shot simulations.
    The slowest convergence with the number of operators added is observed for finite shot simulations for \ac{adapt}. While \ac{dsgn} calculations with a finite amount of shots
    are converged with about 25 parameters to an error of $10^{-9}$ Hartree, the precision for finite shot calculations using
    \ac{adapt} is orders of magnitudes lower, at about $10^{-3}$ Hartree at the same point.\\

    With respect to the resulting Ansatz depth, we observe that Ans{\"a}tze constructed with the order of operators
    resulting from state vector simulations using the \ac{adapt} metric result in the most compact circuits, followed
    by \ac{dsgn} and \ac{enlm}. Ans{\"a}tze constructed with finite shot simulations for \ac{adapt} are the least
    compact.

    The total amount of shots for \ac{adapt} and \ac{algo} are very different. Both \ac{enlm} and \ac{dsgn} converge
    with a total number of shots about two orders of magnitude lower than the number of shots required for finite shot
    simulations using \ac{adapt}. \ac{dsgn} requires fewer shots to obtain a given precision compared to \ac{enlm}.\\

    \subsection{LiH}
    Similar observations as for $\text{H}_{4}$ also hold true for the $\text{LiH}$ calculations presented in Fig.~\ref{fig:lih}, even though the
    overall convergence is slower. There are some other features to be observed in the convergence for this system. For example,
    the finite shot simulations for \ac{adapt} with 100 and 500 shots per operator evaluation showed no sign of convergence and remained on the level of
    the \ac{hf} reference state. The finite shot simulation with 1000 shots per operator evaluation shows early signs of
    convergence but is not able to go much below an energy difference of $10^{-2}$ Hartree. The \ac{adapt} state vector simulation, and all
    simulations for \ac{dsgn} and \ac{enlm} converge to an energy difference of $10^{-3}$ Hartree at roughly the same rate, here the \ac{dsgn}
    convergence flattens out, while the remaining calculations continue to converge at a similar rate. Beyond the addition of roughly 30 parameters
    \ac{dsgn} gets a dramatic increase in precision while the other calculations start flattening out, displaying an unintuitive and seemingly 
    erratic behaviour of convergence. It is notable that \ac{dsgn} converges below the \ac{adapt} state vector simulation.\\

    With respect to Ansatz compactness and the number of shots required, similar conclusions hold true, displaying the same overall tendencies
    as observed for $\text{H}_4$ including the specific features described for the energy evaluation above.
    
    \subsection{Discussion}
    The dramatic difference in the number of shots between \ac{algo} and \ac{adapt} is due to
    the excessive amount of shots necessary to measure the gradients of the operator pool, $\mathcal{A}$, of \ac{adapt}.
    We can write the total amount of shots as iterations times shots for \ac{algo}  whereas for \ac{adapt} it reads iterations
    times shots times $|\mathcal{A}|$. Such a fact also provides another reason for the slow convergence of \ac{adapt}
    when using a finite amount of shots as the evaluation of the gradient in Eq.~\eqref{eq:adaptgrad} is prone to sampling error.
    In contrast, Eqns.~\eqref{eq:weight} and \eqref{eq:altmeasure} for \ac{enlm} and \ac{dsgn} are evaluated on a classical
    computer from states that are generated by the measurement of the energy. However, it remains be noticed that sampling error also
    effects \ac{enlm} and \ac{dsgn} as these methods are dependent on a representation of the
    weights of the determinants in the current wave function $|\Psi^{(k)}\rangle$. For \ac{adapt}, more precise measurements
    of the gradients are required in order to improve convergence, while more precise sampling of the Slater determinants
    (diagonal elements of the Hamiltonian) becomes necessary for \ac{algo}. This is especially important when the electronic
    structure becomes more correlated, i.e., when many determinants are required to describe the chemical system accurately, the
    necessary sampling depth may become a challenge.\\

    It must also be noted that none of the proposed metrics for selecting the next operator is optimal and that there is
    room for improvement. For example, despite being the overall most competitive metric, \ac{dsgn} seems to select some sub-optimal
    operators for $\text{LiH}$ below $10^{-3}$ Hartree, yet it converges at an order of magnitude below the error which the \ac{adapt} state vector
    simulation achieves beyond 60 parameters and \ac{enlm}, which do not exhibit the same behaviour. Additionally, for $\text{LiH}$ with the \ac{enlm} metric
    the finite shot simulations with fewer shots achieve higher precisions indicating that this metric does
    not capture some important correlations in this particular system.\\

    The results shown here suggest that for practical purposes the introduced heuristic metrics are good enough, since they
    converge at a similar rate as the \ac{adapt} state vector simulations using a finite amount of shots. However, the
    systems shown here are rather small and the basis sets are limited. With the two different systems investigated,
    we have observed quite different detailed behaviours of convergence with no clear indication for why the ordering
    behaves so differently with different metrics. A better theoretical understanding of the limits of this method and a more rigorous
    derivation of metrics could make the convergence more robust across many systems and ensure that a similar convergence rate 
    is retained for more complicated molecules and larger basis sets.

    \section{Conclusion} \label{sec:conclusion}
    In this work, we have presented \ac{algo}, a method for selecting operators based on the populations
    of Slater determinants in the wave function. We have introduced two different importance metrics \ac{dsgn} and \ac{enlm}
    and compared them to \ac{adapt} in terms of the convergence to the FCI ground state energy. As was demonstrated, \ac{algo}
    mitigates the significant measurement overhead for \ac{adapt} by utilizing information about the population
    of Slater determinants in the wave function whereas \ac{adapt} must evaluate the expectation value
    of gradient operators. For infinite shots, \ac{adapt} provides the most compact wavefunction in terms
    of CNOT gates but with equal amount of parameters compared to \ac{algo}. For finite shot simulations, \ac{algo} yields more
    compact wave functions with dramatically reduced execution times. Of the two introduced importance metrics \ac{dsgn}
    converged most rapidly and resulted in more compact circuits compared to \ac{enlm}. However, we expect that a more systematic construction of importance metrics may
    improve the performance and eliminate some erratic features seen, e.g., for $\text{LiH}$.
    It remains to be seen how this method performs on real quantum hardware and in combination with other operator pools and other improvements
    available for \ac{adapt}. This will be the topic of future investigations.

    \begin{acknowledgments}
        We thank Niels Kristian Kjærgård Madsen, Mads Bøttger Hansen, Mogens Dalgaard and Stig Elkjær Rasmussen from Kvantify ApS
        and Mads Greisen Højlund, Rasmus Berg Jensen and Ove Christiansen from Aarhus University for fruitful discussions.
    \end{acknowledgments}

    \begin{conflicts}
	NTZ is a co-founder of Kvantify Aps. The authors have filed a provisional patent application covering the 
	method described here.
    \end{conflicts}

    \appendix

    \bibliographystyle{ieeetr}
    \bibliography{fast_vqe.bib}

\end{document}